\documentclass{ws-procs9x6}
\newcommand{\U}{\mathop{\rm {}U}}
\newcommand{\tr}{\mathop{\rm tr}\nolimits}
\newcommand{\Tr}{\mathop{\rm Tr}\nolimits}

\begin{document}

\title{Chiral Anomalies in the Reduced (or Matrix) Model%
\footnote{%
\uppercase{T}alk presented at \uppercase{I}nternational \uppercase{W}orkshop,
``\uppercase{S}trong \uppercase{C}oupling \uppercase{G}auge
\uppercase{T}heories and \uppercase{E}ffective \uppercase{F}ield
\uppercase{T}heories, 10--13 \uppercase{D}ecember 2002,
\uppercase{N}agoya, \uppercase{J}apan.}
\footnote{%
\uppercase{T}his talk is based on a collaboration with
\uppercase{Y}oshio \uppercase{K}ikukawa~[1].}}

\author{Hiroshi Suzuki}

\address{Department of Mathematical Sciences,\\
Ibaraki University, Mito 310-8512, Japan\\
E-mail: hsuzuki@mx.ibaraki.ac.jp}


\maketitle

\abstracts{We show that, with an appropriate choice of a Dirac operator, there
is a remnant of chiral anomalies in the reduced model in which there is no
coordinate dependences of the gauge field. This result is obtained by exploring
a topological nature of chiral anomalies associated to Ginsparg-Wilson-type
lattice Dirac operators.
}

\section{Introduction}
We already had several talks concerning new ideas on a treatment of chiral
fermions in lattice gauge theory in this workshop.\cite{Chiu} The
essence of these new ideas can be summarized in a simple relation which is
called the Ginsparg-Wilson relation. In this talk, I present
another kind of application of these new ideas in lattice gauge theory.

\section{Axial Anomaly with the Ginsparg-Wilson Relation}
The Ginsparg-Wilson relation\cite{Ginsparg:1982bj} of $d$-dimensional lattice
Dirac operator reads
\begin{equation}
   \gamma_{d+1}D+D\gamma_{d+1}=D\gamma_{d+1}D.
\label{one}
\end{equation}
Here $\gamma_{d+1}$ is the $d$-dimensional analogue of $\gamma_5$ and I set the
lattice spacing unity~$a=1$ for notational simplicity. An important consequence
of this relation is a topological property of the axial anomaly, which is
defined by
\begin{equation}
   q(x)\equiv
   \tr\gamma_{d+1}\left[1-{1\over2}D(x,x)\right]
   {\buildrel a\to0\over\rightarrow}
   {1\over2}\partial_\mu\langle
   \overline\psi\gamma_\mu\gamma_{d+1}\psi(x)\rangle.
\end{equation}
From the algebraic relation~(\ref{one}) and the ``$\gamma_5$-hermiticity''
$D^\dagger=\gamma_{d+1}D\gamma_{d+1}$, one can show the ``index
relation''\cite{Hasenfratz:1998ri} on a lattice~$\mathit{\Gamma}$
\begin{equation}
   Q\equiv\sum_{x\in\mathit{\Gamma}}q(x)
   =\Tr\gamma_{d+1}\left(1-{1\over2}D\right)
   =n_+-n_-,
\label{three}
\end{equation}
where $n_\pm$ is a number of zero-modes of $\gamma_{d+1}D$ with
$\pm$~chirality. The {\it integer\/} $Q$ therefore provides a topological
characterization of a gauge field configuration, even on a finite-size lattice.

We take Neuberger's overlap Dirac operator\cite{Neuberger:1998fp} as a definite
example of~$D$:
\begin{equation}
   D=1-{A\over\sqrt{A^\dagger A}},\quad A=1-D_{\rm w},\quad
   D_{\rm w}={1\over2}[\gamma_\mu(\nabla_\mu^*+\nabla_\mu)
   -\nabla_\mu^*\nabla_\mu].
\label{four}
\end{equation}
In this construction, covariant difference operators are defined
by\footnote{$\hat\mu$ denotes a unit vector along the $\mu$-th direction.}
\begin{equation}
   \nabla_\mu=U_\mu(x)T_\mu-1,\quad
   \nabla_\mu^*=1-U_\mu(x-\hat\mu)^\dagger T_\mu^\dagger,
\label{five}
\end{equation}
where $T_\mu$ is the shift operator: $T_\mu\psi(x)=\psi(x+\hat\mu)$.

Now, assuming that the combination~$Q$~(\ref{three}) gives a non-trivial
value~$n_+-n_-\neq0$ (this turns to be actually the case), the index
relation~(\ref{three}) implies that the overlap operator~$D$ cannot be a smooth
function of the gauge field in general: The space of lattice gauge fields is
arc-wise connected and the {\it integer\/} $Q$ cannot ``jump'' unless $D$
becomes singular at certain points (see Fig.~1).

\begin{figure}[ht]
\centerline{\epsfxsize=5cm\epsfbox{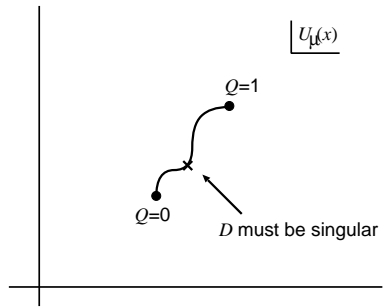}}   
\caption{The overlap Dirac operator~$D$ must be singular at a certain point
on a path from $Q=0$ to $Q=1$.}
\end{figure}

To avoid these singularities, some restriction on gauge field configurations
has to be imposed. For the overlap-Dirac operator, a sufficient condition for
the well-defined-ness of $D$ is known as
``admissibility''\cite{Hernandez:1999et}
\begin{equation}
   \|1-U_\mu(x)U_\nu(x+\hat\mu)U_\mu(x+\hat\nu)^\dagger U_\nu(x)^\dagger\|
   <\epsilon,
\label{six}
\end{equation}
for all plaquettes, where $\|\mathcal{O}\|=%
\sup_{v\neq0}\|\mathcal{O}v\|/\|v\|$. This condition divides (otherwise
arcwise-connected) space of lattice gauge fields into many topological
sectors (see Fig.~2).

\begin{figure}[ht]
\centerline{\epsfxsize=5cm\epsfbox{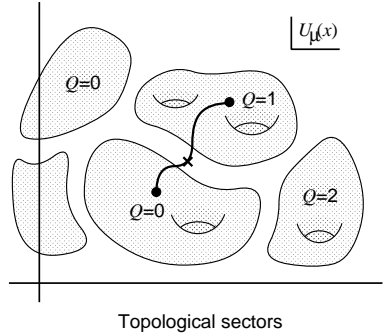}}   
\caption{After imposing the admissibility, topological sectors emerge.}
\end{figure}

With a lattice Dirac operator which obeys the Ginsparg-Wilson relation, the
topological structure of gauge theory naturally emerges in this way. It is
interesting that this picture works even with finite lattice sizes and with
finite lattice spacings, namely for a system of finite degrees of freedom.

Furthermore, when the gauge group is~$\U(1)$, the topological nature of the
axial anomaly allows a non-perturbative cohomological analysis. Using the facts
that $q(x)$ is a gauge invariant local pseudoscalar field which satisfies
$\sum_{x\in\mathit{\Gamma}}\delta q(x)=0$, one can show that the most general
form of~$q(x)$ is given by\cite{Luscher:1999kn}
\begin{eqnarray}
   q(x)&=&\gamma
   \epsilon_{\mu_1\nu_1\cdots\mu_{d/2}\nu_{d/2}}
   F_{\mu_1\nu_1}(x)F_{\mu_2\nu_2}(x+\hat\mu_1+\hat\nu_1)\cdots
\nonumber\\
   &&\quad\times F_{\mu_{d/2}\nu_{d/2}}
   (x+\hat\mu_1+\hat\nu_1+\cdots+\hat\mu_{d/2-1}+\hat\nu_{d/2-1})
\nonumber\\
   &&+\partial_\mu^*k_\mu(x),
\label{seven}
\end{eqnarray}
where $F_{\mu\nu}(x)={1\over i}\ln
U_\mu(x)U_\nu(x+\hat\mu)U_\mu(x+\hat\nu)^\dagger U_\nu(x)^\dagger$
and $k_\mu(x)$ is a certain gauge invariant local periodic current
on~$\mathit{\Gamma}$. This is the first example (to my knowledge) that one can
see an explicit structure of the axial anomaly in a system with finite UV and
IR cutoffs. This non-perturbative information was fully utilized in a
manifestly gauge invariant lattice formulation of $\U(1)$ chiral gauge
theories.\cite{Luscher:1999du} The coefficient~$\gamma$ can be obtained by a
matching with the classical continuum limit
as~$\gamma=(-1)^{d/2}/[(4\pi)^{d/2}(d/2)!]$.\cite{Kikukawa:1998pd}

The general argument~(\ref{three}) tells that $Q$ is an integer. One can see
this fact more directly when the gauge group is $\U(1)$. In this case, gauge
fields which satisfies the admissibility~(\ref{six}) can be completely
classified as\cite{Luscher:1999du}
\begin{equation}
   U_\mu(x)=\Lambda(x)V_\mu^{[m]}(x)U_\mu^{[w]}(x)
   e^{iA_\mu^{\rm T}}(x)\Lambda(x+\hat\mu)^{-1},
\end{equation}
where $\Lambda(x)$ is the gauge degrees of freedom and $U_\mu^{[w]}(x)$ carries
the Polyakov line, $\prod_{s=0}^{L-1}U_\mu^{[w]}(s\hat\mu)=
w_\mu\in\U(1)$.\footnote{$L$ is a size of the lattice.}
The part~$V_\mu^{[m]}(x)$ has a constant field strength
$F_{\mu\nu}(x)=2\pi m_{\mu\nu}/L^2$, where $m_{\mu\nu}\in{\mathbb Z}$,
and $A_\mu^{\rm T}(x)$ represents gauge invariant local fluctuations. Then, by
using eqs.~(\ref{three}) and~(\ref{seven}), we have
\begin{equation}
   Q={(-1)^{d/2}\over2^{d/2}(d/2)!}\,
   \epsilon_{\mu_1\nu_1\cdots\mu_{d/2}\nu_{d/2}}
   m_{\mu_1\nu_1}m_{\mu_2\nu_2}\cdots m_{\mu_{d/2}\nu_{d/2}},
\label{ten}
\end{equation}
for $\U(1)$ theories; this is manifestly an integer.

Here we present an application of above ideas to the reduced model.

\section{Reduced model and the $\U(1)$ embedding}

The $\U(N)$ reduced model is defined by a zero-volume limit of $\U(N)$ lattice
gauge theory, $U_\mu(x)\in\U(N)\to U_\mu\in\U(N)$. This kind of reduction
appears in the Eguchi-Kawai reduction of the large $N$ QCD\cite{Eguchi:1982nm}
and in the compact version of IKKT IIB matrix model.\cite{Kitsunezaki:1997iu}
Here we concentrate on the fermion sector in the reduced model:
$\psi(x)\to\psi\in\hbox{fundamental rep.\ of $\U(N)$}$.
According to above prescriptions, the covariant derivative for the
reduced fermion field would be read as\footnote{This corresponds to a
``naive'' prescription. Our argument in what follows is applicable to
the quenched reduced model\cite{Levine:1982uz} as well.}
\begin{equation}
   \nabla_\mu=U_\mu(x)T_\mu-1\to U_\mu-1.
\label{twelve}
\end{equation}
In the reduced model, one would expect that there is no anomaly as
$\tr F\cdots F(x)\to \tr A\cdots A\equiv0$. We will show that in an appropriate
framework there exists some remnant of chiral anomalies even in
this zero-dimensional field theory.

To show this, we first assume~$N=L^d$, where $L$ is an integer. Then we can
identify an index of the fundamental representation~$n$ and a site~$x$ on a
lattice~$\mathit{\Gamma}$ of the size~$L$, $x=(x_1,\ldots,x_d)$, by
$n(x)=1+x_d+Lx_{d-1}+\cdots+L^{d-1}x_1$.
With this identification, the $N\times N$
matrix~$T_\mu=1\otimes\cdots\otimes1\otimes X
\otimes1\otimes\cdots1$, where the factor $X$
\begin{equation}
   X=\pmatrix{0&1&&\cr
              & \ddots & \ddots &\cr
              &        & \ddots &1\cr
              1&&&0\cr},
\end{equation}
appears in the $\mu$-th entry, realizes the shift on the lattice (while
preserving the periodic boundary condition) as
$T_\mu f_{n(x)}=f_{n(x+\hat\mu)}$.

So next we assume that the reduced gauge field has the following particular
form
\begin{equation}
   U_\mu=u_\mu T_\mu,
\label{fourteen}
\end{equation}
where $u_\mu$ is a {\it diagonal\/} matrix
\begin{equation}
   u_\mu=\pmatrix{(u_\mu)_1&&\cr
              & \ddots &\cr
              &&(u_\mu)_N\cr}.
\end{equation}
Note $(u_\mu)_n\in\U(1)$. Then the covariant derivative in the reduced model
takes the form
\begin{equation}
   \nabla_\mu=U_\mu-1=u_\mu T_\mu-1.
\end{equation}
By comparing this with eq.~(\ref{five}), we realize that {\it the fermion
sector of the reduced model with $U_\mu=u_\mu T_\mu$ is completely equivalent
to that of the conventional $\U(1)$ lattice gauge theory}. The $\U(1)$ gauge
field in the latter is diagonal elements of the matrix $u_\mu$. In this sense,
we call eq.~(\ref{fourteen}) $\U(1)$ embedding.

An interesting property of the $\U(1)$ embedding is that the plaquette is
identical for both pictures:
\begin{eqnarray}
   U_\mu U_\nu U_\mu^\dagger U_\nu^\dagger
   &=&u_\mu(T_\mu u_\nu T_\mu^\dagger)
   (T_\nu u_\mu^\dagger T_\nu^\dagger)
   u_\nu^\dagger
\nonumber\\
   &=&u_\mu(x)u_\nu(x+\hat\mu)u_\mu(x+\hat\nu)^*u_\nu(x)^*,
\label{seventeen}
\end{eqnarray}
due to the relation $(T_\mu fT_\mu^\dagger)_{m(x)n(y)}
=f_{m(x+\hat\mu)n(y+\hat\mu)}$ which holds for a diagonal matrix~$f$.
Also the trace on the side of the reduced model is simply written by a
lattice summation, $\tr f=\sum_{x\in\mathit{\Gamma}}f(x,x)$. In this way,
we can switch between matrix- and lattice-pictures.

\section{Vector-like reduced model and the topological charge}

Following the proposal of Kiskis, Narayanan and Neuberger,\cite{Kiskis:2002gr}
we use the overlap-Dirac operator in the reduced model. It is defined by the
construction~(\ref{four}) with the substitution~(\ref{twelve}) and
$\nabla_\mu^*=1-U_\mu^\dagger$. As they demonstrated for $d=2$ and $4$ by
using a somewhat different idea, this framework provides a well-defined
topological charge for reduced gauge fields. To make the overlap-Dirac operator
well-defined, one imposes the admissibility
$\|1-U_\mu U_\nu U_\mu^\dagger U_\nu^\dagger\|<\epsilon$ on reduced gauge
fields. Then the axial anomaly (or the topological charge) in the reduced
model, $Q=\tr\gamma_{d+1}(1-D/2)$, provides a well-defined topological
characterization of the reduced gauge field. Recall Fig.~2.

By further using the $\U(1)$ embedding~(\ref{fourteen}), we can use the
above matrix-lattice correspondence. The crucial point is that the
admissibility is common for both pictures as eq.~(\ref{seventeen}) shows.
So we can literally copy results in Sec.~2! In this way, we immediately
find that $Q$ in the reduced model is given by eq.~(\ref{ten}).
Note that the integers $m_{\mu\nu}$ this time parameterize a form of
matrices~$U_\mu$. See ref.~[1]. By this way, we see that there exist
reduced gauge fields which have non-trivial topological charges.

\section{Chiral gauge reduced model and an obstruction}

By using Ginsparg-Wilson-type Dirac operator, one can formulate chiral gauge
reduced model, along the line of ref.~[8]. This formulation is equivalent to
the overlap~[14]. A complexity of this formulation is, however, the fermion
integration measure has an ambiguity in its phase and one has to fix this
ambiguity somehow. After imposing the admissibility, the space of gauge fields
may have a complicated topology (see Fig.~2) and then it is not obvious whether
the phase can be chosen as a single-valued function on this space.
This problem can be formulated in terms of a $\U(1)$ fiber bundle associated
to a phase of the fermion measure.\cite{Neuberger:1998xn,Luscher:1999du} If and
only if this bundle is trivial, one can define a single-valued expectation
value in the fermion sector. One of measures of a non-triviality of the bundle
is the first Chern number (the monopole charge) $\mathcal{I}$ defined for
closed 2~dimensional surfaces in the space of admissible gauge fields. If we
have $\mathcal{I}\neq0$ for a certain surface, the bundle cannot be trivial
and {\it a Weyl fermion cannot be consistently formulated}.

By using the $\U(1)$ embedding and a cohomological analysis of
L\"uscher's topological field in $d+2$~dim.,\cite{Luscher:1999du} we found
there exists a 2-torus such that $\mathcal{I}\neq0$. Space does not permit a
detailed presentation. See ref.~[1]. This shows that a Weyl fermion in the
fundamental rep.\ of $\U(N)$ in the reduced model cannot be consistently
formulated within this framework. We regard this as a remnant of the gauge
anomaly of the original gauge theory. A generalization of this result to other
gauge-group representations is under study.

\end{document}